\documentclass[aps,pra,10pt, showpacs,english,keywords,keywords,amsmath,amssymb,twocolumn,superscriptaddress]{revtex4-1}
\usepackage[T1]{fontenc}
\usepackage[latin1]{inputenc}
\usepackage{babel}
\usepackage{graphicx}
\usepackage{color}
\usepackage{bm}
\usepackage{longtable}
\usepackage{amsmath}
\usepackage{amsfonts}
\usepackage{amssymb}
\usepackage{hyperref}

\begin{document}
\title{ Single phase and correlated phase estimation with multi-photon annihilated squeezed vacuum states: An energy balancing scenario}
\author{N.~Samantaray}
\email{Email: ns17363@bristol.ac.uk}
\affiliation{Quantum Engineering Technology Labs, H. H. Wills Physics Laboratory and Department of Electrical and Electronic Engineering, University of Bristol, BS8 1FD, UK}

\author{I.~Ruo-Berchera}
\affiliation{INRIM, Strada delle Cacce 91, I-10135 Torino, Italy}
\author{I.P.~Degiovanni}
\affiliation{INRIM, Strada delle Cacce 91, I-10135 Torino, Italy}
\affiliation{INFN, via P. Giuria 1, I-10125 Torino, Italy}

\begin{abstract}
In the last years, several works have demonstrated the advantage of photon subtracted Gaussian states for various quantum optics and information protocols. In most of these works, it was not clearly investigated the relation between the advantages and the usual increasing energy of the quantum state related to photon subtraction.  In this paper, we study the performance of an
interferometer injected with multi photon annihilated squeezed vacuum states mixed with coherent states for both single and correlated phase estimation. For single phase estimation, albeit the use of multi-photon annihilated squeezed vacuum states at low mean photons per mode provide advantage compared to classical strategy, when the total input energies is held fixed, the advantage due to photon subtraction is completely lost. However, for the correlated case in analogous scenario, some advantage appears to come from both the energy rise and improvement in photon statistics. In particular quantum enhanced sensitivity with photon subtracted states appears more robust to losses, showing an advantage of about 30$\%$ with respect to squeezed vacuum state in case of realistic value of the detection efficiency.
\end{abstract}
\maketitle

\section{Introduction}
Non-Gaussian  states have been recognized as a valuable resources for many quantum information processing protocols \cite{Genoni:2010}, for example to enhance the fidelity of quantum teleportation \cite{Olivares:2003,Optarny:2000, Cochrane:2002, Dell'Anno:2010}, improving  secret key rate in quantum key distribution \cite{Borelli:2016, Liao:2018}, and in quantum cloning of coherent states \cite{Cerf:2005}. These exotic states are required in Gaussian entanglement distillation \cite{Eisert:2002, Cirac:2002, Fiurasek:2002}, error correction \cite{Niset:2009}, noiseless amplification \cite{Ferreyrol:2010, Xiang:2010}, and fundamental loop-hole free Bell tests in continuous variables \cite{Banaszek:1998, Nha:2004}. More recently, resource theories quantifying the importance of Wigner negativity and non-Gaussianity for continuous variable quantum computation have been reported \cite{Takagi:2018, Albarelli:2018}. Because of higher potential distinguishability of non-Gaussian states from their original Gaussian states, they have been proven useful for more precise parameter estimation in quantum optics \cite{Genoni:2009, raul:2012}.
Photon addition or photon subtraction can transform a Gaussian in a non-Gaussian state \cite{kim:2008, kim1:2008, Bellini:2007}. Tara and Agarwal \cite{tara:1990} were the first to propose the transformation of a classical like coherent state into a non-classical state through photon addition and this operation was experimentally implemented for the first time on coherent and thermal state in \cite{ Bellini:2004,Bellini1:2007}. Furthermore, photon addition and subtraction have been reported to enhance the entanglement in two mode squeezed vacuum state (TSV) \cite{Carlos:2012,Tim:2013,ourjoumtsev:2007}. It is well known that each mode of the TSV has super-poissonian photon statistics. In the experimental work\cite{chekova:2016}, it has been reported that multi-photon subtraction makes the TSV less noisy and helps in shifting  the most probable distribution to higher mean photon number, thereby it increases the mean energy of the resulting state. In the last years, photon subtracted TSV states have been theoretically investigated for other applications, e.g. in \cite{Zhang:2014} these states were proved to be advantageous with respect to TSVs for target detection in noisy environment, a scheme dubbed "quantum illumination" \cite{Lloyd:2008,Lopaeva:2013}. Their advantage has also been demonstrated in single interferometry with parity measurements\cite{raul:2012,Birritella:2014,yoa:2016}. 

\par Since the seminal Caves work \cite{Caves:1981}, it is well known that single mode squeezed vacuum (SSV) mixed with intense coherent state provides substantial advantage in practical phase estimation and very recently that scheme has been applied by LIGO and VIRGO collaborations to further improve the sensitivity of gravitational waves detectors \cite{Tse:2019, Acernese:2019}. It has also been shown that squeezed vacuum mixed with a intense coherent beam allow to approach the optimal sensitivity achievable by a linear interferometer operated with large photon number and non negligible losses \cite{Rafal:2013}. In that context, single photon subtracted SSV brings advantage in phase estimation and allow reaching Heisenberg's limit (HL) \cite{olivares:2016}. In that case \cite{olivares:2016}, the mean energy increasing of the photon subtracted state has been compensated by a reduction of the coherent beam energy in order to keep constant the total input photon number. Photon subtracted SSV mixed with coherent states lead to improved phase shift sensitivity in parity measurement \cite{Birritella:2014}. In this work, because of non-linear increase of mean number of photons with the number of subtracted photons, the total average number of photons is fixed by choosing fixed squeezing parameter and increased coherent energy. This approach is little different than the energy balancing scenario reported in \cite{olivares:2016}.

%Squeezed vacuum affinity for any arbitrary state is the maximum overlap between the state and the squeezed vacuum state. Thus larger is the non-Gaussianity, more is the squeezed vacuum affinity in terms of overlap and higher is the deportation fidelity when compared for a given squeezed parameter accounting higher energies of subtracted states \cite{Dell'Anno:2007}.

However, in most of the literature with some exception that we will point out later \cite{raul:2012,Dell'Anno:2007}, it was not clear whether the advantages come from energy shifts of the photon subtracted states, or from their potentially improved photon noise properties. In fact, most of the time, the comparison between the performance of Gaussian quantum states and the correspondent photon subtracted states has been considered at fixed squeezed parameter, that means generally that the energy put into the quantum resource is not fixed. Photon subtraction is a complex operation that can be experimentally realized probabilistically or with low efficiency \cite{Marek:2018}. So, if the advantage comes mainly from the increased energy, it can practically more convenient to increase the squeezing parameter of the Gaussian state, rather than performing photon subtraction. This is the reason why we consider of great importance understanding if the advantage relies solely on increase of the energy, or there are more fundamental reasons that justify operation such as photon subtraction. Answering to this question is the main motivation of this paper. 

Specifically, given the importance for practical interferometry \cite{Rafal:2013}, here we study in detail multi-photon subtracted single mode and two mode squeezed vacuum state for single phase and correlated phase estimation respectively by combining them with coherent states on a beam splitters. On one side, we show that a multi-photon subtracted (one-) two-mode squeezed states is formally equivalent to a state obtained by a (one-) two- mode squeezing operator applied to a certain class of finite superposition states in the photon number basis. This class of states have been investigated earlier \cite{Wodkiewicz:1987,figurny:1993} and they show quadrature squeezing their-self. One could expect that this initial squeezing could bring benefit in phase estimation.  We have therefore investigated this possibility.

%On the other side the equivalence between photon subtracted squeezed states and squeezing of superposition states, allows to devise alternative strategy for their generation, usually based on probabilistic post-selection conditioned by photon detection events.  This is another important results shown in our work. As an example, a one photon subtracted squeezed state can be generated by seeding a squeezer (non-linear parametric amplifier) by a single photon state. More in general, the number of elements of the superposition state  which are necessary for seeding the non-linear process is in simple relation with  the order of photon subtraction.
% 
In order to properly understand the origin of the improvement in phase measurement uncertainties if any, we think that the proper procedure requires that the total energy should be fixed by balancing the energies of the subtracted and un-subtracted states keeping the coherent pump energy constant. We will consider this energy balancing condition for both single and correlated phase estimations. Similar analysis has been done in \cite{raul:2012}, where an advantage in phase estimation with parity measurement at fixed energy has been reported, but that scheme does not involve the mixing with a coherent state, and parity measurement is quite far from realistic applications. In \cite{Dell'Anno:2007}, while a precise comparison with two-mode squeezed states with the same energy has not been carried out they showed that the larger is the "affinity" of a non-Gaussian state with a two-mode squeezed vacuum (with larger energy), the larger is the teleportation fidelity. 

%In general we find that energy increasing of the states  intrinsic to the photon subtraction operations is in most cases the origin of the advantage of the photon subtracted states. However, in case of the specific scheme of correlation measurement among two interferometer by exploiting bipartite correlated states such as TSV, the advantage of the symmetric photon subtraction cannot be explained only by energy shift.

This paper is organized as follows. In Sec. II, we will introduce multi-photon annihilated single mode squeezed vacuum (PASSV), discussing their properties and their usefulness, for single phase estimation by the conventional measurement strategy \ref{Single phase estimation with PASSV states} and in the more general framework of the fisher information \ref{Quantum fisher information}. In Sec. \ref{ SPATSV states}, we will describe multi-photon symmetrically annihilated two mode squeezed vacuum (SPATSV) state. In particular, in section \ref{Squeezing and photon statistics of the SPATSV state}, we will analyze squeezing and photon statistical properties of SPATSV. In section \ref{correlated phase estimation} we study the problem of correlated phase estimation. We present results up to four and three photons subtraction for single and correlated phase estimations respectively. Finally, we will summarize and discuss the main results in Sec. \ref{Summary and conclusions}.

\section{PASSVS}\label{ PASSVS}

PASSV states are defined as:
\begin{equation}
\vert\Psi^{(m)}_{PASSV}\rangle =N^{m}_{-}(\lambda)\hat{a}^{m}\hat{S}(\lambda e^{i\chi})\vert 0\rangle,
\end{equation}
where $\hat{S}(r e^{i\chi})= e^{r e^{i\chi} a^{2} - h.c.}$ is the single mode squeezing operator, $r$ being the squeezing parameter and $\chi$ is the squeezing angle. The squeezing operator applied to the vacuum state originates SSV with energy (mean number of photon) equal to with $\lambda=\sinh^{2} r$. The number of subtracted photon is $m$, obtained by $m$ consecutive action of the annihilation operator $\hat{a}$. Since the photon subtraction is not a unitary operation, it is necessary to introduce the normalization constant $N^{m}_{-}(\lambda)$. Its explicit form is $N^{m}_{-}(\lambda)=m!(-i\sqrt{\lambda})^{m}P_{m}(i\sqrt{\lambda})$ \cite{nigam:2017}, with $P_{m}$ being the $m$-th order Legendre's polynomial. A known effect of the photon subtraction is the increasing of the mean energy of the state. This is intuitively explained because it is more likely to subtract a photon from a highly populated state corresponding to a selection of the more energetic components of the state. For example, the mean photon number of PASSV state $\mathcal{N}_{m}(\lambda)$, for $m=0-3$ which correspond to zero, one, two and three photon subtraction from the SSV state, becomes $\mathcal{N}_{0}=\lambda$, $\mathcal{N}_{1}=3\lambda+1$, $\mathcal{N}_{2}=3\lambda(3+5\lambda)/(1+3\lambda)$, and $\mathcal{N}_{3}=(3+30\lambda+35\lambda^{2})/(3+5\lambda)$ respectively.
   
We have found a new representation for PASSV states (exploiting integration within an ordered product (IWOP) technique \cite{Fan:2003}) which is equivalent to seeding squeezing operator with photon number superposition state $\vert\Theta^{s}_{m}(\lambda,\chi)\rangle$ in input as follows:
\begin{equation}
 \vert\Psi^{(m)}_{PASSV}\rangle= \hat{S}\vert\Theta^{s}_{m}(\lambda,\chi)\rangle,\nonumber
 \end{equation}
 \begin{eqnarray}\label{passv}
 && \vert\Theta^{s}_{m}(\lambda,\chi)\rangle = N^{m}_{-}(\lambda)m!\left(e^{i\chi}\sqrt{\lambda}\right)^{m}\times\\ && \nonumber  \sum_{l=0}^{[m/2]}\frac{1}{l!\sqrt{(m-2l)!}}\left(\frac{e^{-i\chi}}{2}\sqrt{\frac{1+\lambda}{\lambda}}\right)^{l}\vert m-2l\rangle,
 \end{eqnarray}
where the upper bound of the summation $[m/2]$ stand for the integer part of $m/2$. Without any loss of generality, hereinafter we set the squeezing angle to $\chi=0$. 
For $m=0$, the input simplifies to vacuum state as expected. For $m=1$, it becomes a single photon state as reported in \cite{olivares:2016}. Note that, for other values of $m$, it becomes a $(m+1)/2$ components superposition state for odd $m$, and a $(m+2)/2$ components superposition state for even $m$. The energy increasing with $m$ of the PASSV state can now be understood in terms of increasing of the mean number of photons of the seeding states.   
\begin{figure}[thb]
	\centering
	\includegraphics[width=7 cm]{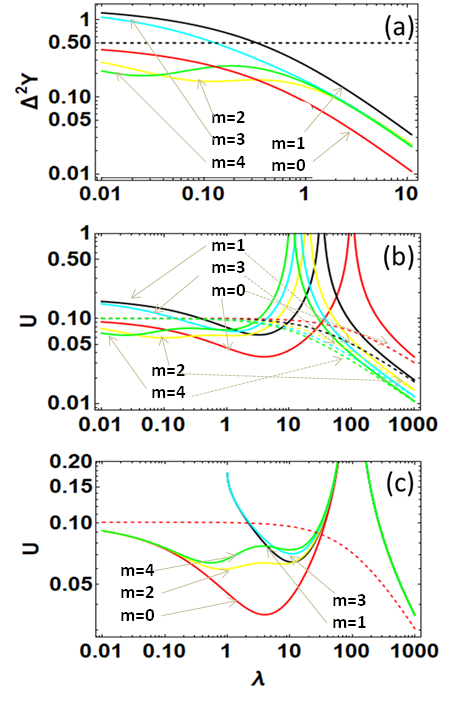}
	\caption{(Color online) Quadrature squeezing of PASSV states and phase measurement uncertainty at working point $\phi=\pi/2$ with $\mu=100$, $\eta=0.98$, and $\psi=0$ for different number of photon subtraction $m$: $m=0$ (solid red line), $m=1$ (solid black line), $m=2$ ( solid yellow line), $m=3$ (solid cyan line), and $m=4$ (solid green line). Dashed lines represents the coherent state: (a.) quadrature squeezing, (b.) phase uncertainty, and (c.) phase measurement as per balancing condition.}\label{squeezingS}
\end{figure}
The states $\vert\Theta^{s}_{m}(\lambda,\chi)\rangle$  are known to show quadrature squeezing \cite{Wodkiewicz:1987,figurny:1993}, even though they can not be obtained by any unitary transformation on vacuum state, like standard squeezed state. However, these states do not always have lower quadrature noise compared to vacuum state. In particular, for $m=1$, the state $\vert\Theta^{s}_{1}(\lambda,\chi)\rangle$ is a single photon state having more quadrature noise than the vacuum state. We checked for subsequent odd values of $m$, although the quadrature noise of the seeding states decrease with respect to single photon state, its value still remain above the vacuum noise. This can be appreciated in Fig \ref{squeezingS}a, when considering small value of the squeezing parameter that actually means $\hat{S}\approx \mathbb{I}$, with $\mathbb{I}$ the identity operator. In fact, in  Fig \ref{squeezingS} we plot the variance of the quadrature  $\hat{Y}=\left(\hat{a}-\hat{a}^{\dagger}\right)/i\sqrt{2}$ of the PASSV states in Eq. (\ref{passv}). In general, we observe that for odd values of $m$ the quadrature noise of the PASSV state is worse than the one of the SSV (corresponding to PASSV with $m=0$), while for even values of $m$ the quadrature noise is better than SSV for low value of $\lambda$. Detectors are not ideal in realistic scenarios. The effect of a non-unit quantum efficiency $\eta$ can be modelled as the evolution of the input field  passing through a beam splitter (BS) with transmission equal to $\eta$, while the other free port of the BS is in vacuum state\cite{Mandel:1995}. This approach has been used throughout the paper to account for the optical/detection losses.

In the next subsection, we will discuss the performance of PASSV states in phase estimation in connection to the quadrature squeezing.

\subsection{Single phase estimation with PASSV states} \label{Single phase estimation with PASSV states}
\begin{figure}[thb]
	\centering
	\includegraphics[width=9 cm]{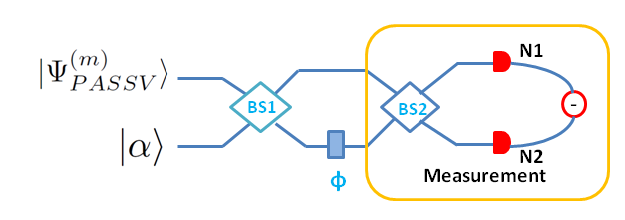}
	\caption{Schematic of mixing single mode squeezed vacuum and coherent state in MachZehnder interferometer for phase estimation $\phi$.}\label{mzi}
\end{figure}

Let us consider the Mach Zehnder interferometer (MZI) sketched in Fig.\ref{mzi}, where one ports of the first beam splitter is injected with coherent light and the other port with a PASSV state. Thus, the total input state is $\vert\Psi^{(m)}_{PASSV}\rangle_{1}\otimes\vert\alpha\rangle_{2}$, with $\alpha=\vert\alpha\vert e^{i\psi}$, where $\vert\alpha\vert$ ($\mu=\vert\alpha\vert^{2}$) and $\psi$ are the amplitude, the mean photon number and the phase of the coherent pump respectively. 

%Our principal objective here is to check if there is any improvement in phase measurement by de-Gaussifying SVS through multi-photon annihilation.

The uncertainty in measuring the phase $\phi$, in the configuration of Fig.\ref{mzi}, is expressed by 
\begin{equation}
U(\phi)=\frac{\sqrt{\Delta^{2}\hat{o}}}{|\frac{\partial \hat{o}}{\partial\phi}|}\label{UUnc},
\end{equation}

where $\hat{o}$ is the photon number difference operator at the output port of the interferometer and $\Delta^{2}\hat{o}$ is its variance. For zero-mean quadrature field such as SSV, $\langle\hat{o}\rangle=(\mu -\lambda) \cos(\phi)$. For SSV, it can be shown that the lowest uncertainty is reached for $\phi=\pi/2$ and in the limit of  $\lambda\gg\mu$, the uncertainty is shot-noise limited, scaling as $\lambda^{-1/2}$.  Whereas in case of $\lambda\ll\mu$, the uncertainty is $U(\phi)=\left(\Delta^{2}X_{\theta=\psi+\frac{\pi}{2}}/\mu\right)^{1/2}$, proportional to the noise of the rotated quadrature $\hat{X}_{\theta=\psi+\frac{\pi}{2}}=\left(\hat{a}e^{-i\theta}+\hat{a}^{\dagger}e^{i\theta}\right)/\sqrt{2}$. In our case and for the choice of $\psi=0$, the sub-shot noise sensitivity is related to the squeezing of the $X_{\frac{\pi}{2}}\equiv Y$ quadrature. For the sake of completeness we mention that for a more specific  repartition of the total energy ($\mu+\lambda$) between squeezing and coherent input states, a more efficient scaling of the uncertainty with can be in principle achieved ($\propto (\mu+\lambda)^{-3/4}$) \cite{Demkowicz:2015}, and different more sophisticated detection scheme could allow to approach the Heisenberg limit (ideal decoherence-free scenario) \cite{Pezze:2008}

\par We have derived analytically the uncertainty on the phase estimation according to Eq. \ref{UUnc}, when PASSV states are injected. The results are shown graphically in Fig\ref{squeezingS} b, compared with the shot-noise limit (SNL) at equivalent total energy (dotted lines). The last is obtained considering the performance of a coherent state with mean number of photon equal to the sum of $\mu$ and the mean number of photons of the PASSV state which varies with $m$. It is easy to check that for PASSV the uncertainty always approach asymptotically the SNL when  $\lambda\gg\mu$. For $\lambda\ll\mu$, the uncertainty is basically determined by the variance of the $Y$ quadrature, reported in Fig\ref{squeezingS} a, as expected. Indeed the advantage over the SNL is present only in the region of quadrature squeezing, and PASSV($m>0$) performs better than SSV only for even $m$. However we are going to show that this apparent improvement is only due to the energy increasing of the state due to photon subtraction. For that purpose, we have renormalized the energy of the initial SSV state before the photon subtraction, so that the mean number of photon of the subtracted states ($m=0-4$) are all equal to $\lambda$.  In this way, also the total input energies to the interferometer is fixed to $N_{tot}=\mu+\lambda$. With the energy balancing, SSV outperforms PASSV regardless of the values of $\lambda$, as presented in Fig\ref{squeezingS}c. 
\begin{figure}[thb]
	\centering
	\includegraphics[width=7 cm]{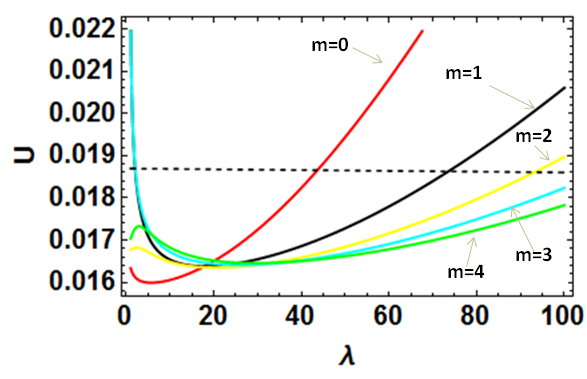}
	\caption{(Color online) Phase measurement uncertainty in energy balancing scenario for $\phi \neq \pi/2\approx \pi/2-1$ with $\mu=10000$, and $\eta=0.98$ for different number of photon subtraction $m$: $m=0$ (solid red line), $m=1$ (solid black line), $m=2$ ( solid yellow line), $m=3$ (solid cyan line), and $m=4$ (solid green line). Dotted line is the classical strategy.}\label{anyangle}
\end{figure}
The diverging behaviour of uncertainty in Fig.\ref{squeezingS} b-c for $\lambda=100$ comes from the singularity in the denominator of Eq.\ref{UUnc} that is when the mean number of photons of the PASSV state equals the one of the coherent state, i.e.  $N_{m}(\lambda)=\mu$. Because of energy increment as a result of photon subtraction, relatively lower value of $\lambda$ is required for  fulfilling the singularity condition as evident from Fig\ref{squeezingS} b.
Incidentally, we have observed that far from the optimal working point $\pi/2$, PASSV can still provide some advantage, even under energy balancing conditions as shown in Fig. \ref{anyangle}, anyway without surpassing the sensitivity obtained by SSV in the optimal point. Typically, this happens from value of $\lambda$ in a middle range (namely from $\mu$/ 100 <$ \lambda $ < $\mu/ 10$). 

Next we will see the advantage if any in QFI perspective.

\subsection{Quantum fisher information}\label{Quantum fisher information}

Quantum Fisher information (QFI), $F_{Q}$, can be used to identify the lower uncertainty attainable in a parameter estimation problem according to the expression

\begin{equation}\label{ineq}
U(\phi)\geq\frac{1}{\sqrt{F_{Q}(\phi)}}.
\end{equation}
For class of pure states \cite{Demkowicz:2015}, QFI takes the following compact form
\begin{equation}\label{qfi}
 F_{Q}(\phi)=4\langle(\Delta \hat{H})^{2}\rangle_{\vert\psi\rangle_{1,2}},
\end{equation}
where  $\hat{H}$ is the generator of the unitary transformation associated with the parameter $\phi$, i.e $\hat{U}(\phi)=e^{i\hat{H}\phi}$ and $\vert\psi\rangle_{1,2}$ being the input states injected into the interferometer. In the case of the MZI of Fig.\ref{mzi},  the generator is the photon number operator $\hat{n}_{3}=\hat{a}^{\dagger}_{3}\hat{a}_{3}$ where $\hat{a}_{3}=(\hat{a}_{1}+\hat{a}_{2})/\sqrt{2}$. 
As per Eq. (\ref{qfi}), we shall evaluate QFI by considering PASSVs and coherent state injected into the interferometer.
The complete expressions of QFI are cumbersome: graphical representation can help to understand its peculiar features. 
\begin{figure}[thb]
\centering
\includegraphics[width=6 cm]{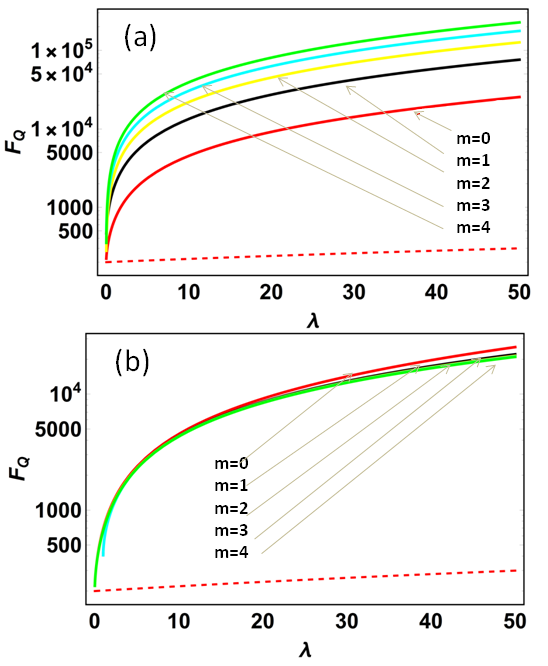}
\caption{(Color online) Quantum fisher information versus mean number of photon $\lambda$ with $\mu=100$ for different number of photon subtraction $m$: $m=0$ (solid red line), $m=1$ (solid black line), $m=2$ ( solid yellow line), $m=3$ (solid cyan line), and $m=4$ (solid green line). Dotted line represents classical bound obtained when only a coherent state with the same total energy ($\lambda+\mu$) is in input: (a.) without energy balance (b.) balanced condition.}\label{fqfi}
\end{figure}
Specifically, Fig.\ref{fqfi}(a) shows a general increasing of the QFI for PASSVs at the increasing of $m$, for both low and high values of $\lambda$, i.e. photon subtraction is always advantageous with respect to best classical strategies  (dotted line). This advantage in QFI due to photon subtraction anyway can be due to the increase of the mean number of photons of the input state. Indeed, for energy balancing condition, the advantage is completely lost as evident from Fig.\ref{fqfi}(b) and also from the expression of QFI in the limit $N_{tot}\rightarrow \infty $ (at a finite fixed coherent energy)  reported here:
\begin{eqnarray}
F_{Q(m=0)}&=& 2N^{2}_{Tot}, \\ F_{Q(m=1)}&=& \nonumber \frac{2N^{2}_{Tot}}{3}, \\ F_{Q(m=2)}&=& \nonumber \frac{2N^{2}_{Tot}}{5},\\ F_{Q(m=3)}&=& \nonumber \frac{2N^{2}_{Tot}}{7}\\ F_{Q(m=4)}&=& \nonumber \frac{2N^{2}_{Tot}}{9}.
\end{eqnarray}
This confirms that the advantage in phase parameter estimation in a MZI provided by photon subtracted of squeezed state is exclusively due to the increasing in the energy of the field. Using a simple SSV state with the same energy provides similar sensitivity.

\section{ SPATSV states}\label{ SPATSV states}
\begin{figure}[htb]
\centering
\includegraphics[width=8cm]{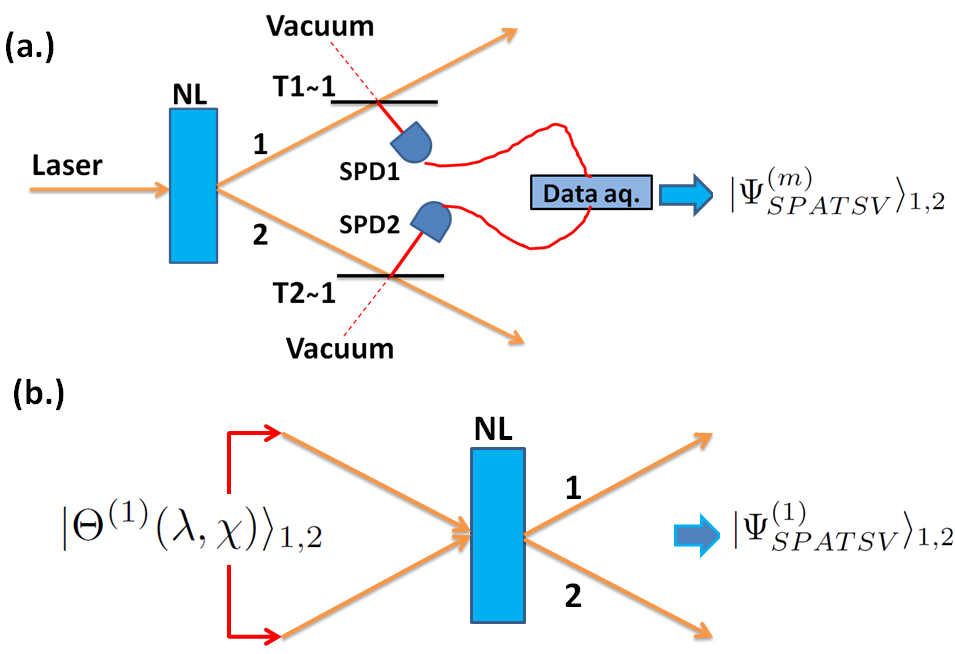}
\caption{Generation scheme for SPATSV state with symmetric photon subtraction of one photon $m=1$:(a) Probabilistic process by two beam splitter of high transmittance $T_{1}=T_{2}\approx1$ placed in two arms of the PDC source. Simultaneous clicks on the two single photon detectors confirms the generation of SPATSV. (b) Alternative approach to the generation of SPATSV consists in injecting seeding state of the general form in Eq. (\ref{sup2}) into the non-linear crystal (NL), in particular the one reported in Eq. (\ref{pho_m=1}) for the case $m=1$.}\label{generation}
\end{figure}

Starting from the definition of the TSV as the two mode squeezing operator $\hat{S}_{1,2}\left(r_{12} e^{i\chi}\right)=e^{r_{12} e^{i\chi} a_{1} a_{2}- h.c.}$ applied to the vacuum, the  SPATSV can be obtained by the non Hermitian operation represented as 
\begin{eqnarray}\label{SPATSV}
 \vert\Psi^{(m)}_{SPATSV}\rangle_{1,2}= \textbf{N}^{-}_{m}\left(\lambda\right)(\hat{a_{1}})^{m}(\hat{a_{2}})^{m} \hat{S}_{1,2}\vert 0,0\rangle_{1,2},
\end{eqnarray}
 where $\textbf{N}^{-}_{m}$ is the normalization constant, $\lambda=\sinh^{2} r_{12}$ is the mean energy (mean photon number) per mode for the TSV, $\chi$ is squeezing angle and $m$ is the number of subtracted photons. It is possible to express the state in Fock basis as follows

\begin{equation}\label{numex}
\vert\Psi^{(m)}_{SPATSV}\rangle=\frac{\textbf{N}^{-}_{m}\left(\lambda\right)}{\sqrt{1+\lambda}}\sum_{n=0}^{\infty}\left(\frac{\lambda e^{i\chi}}{1+\lambda}\right)^{\frac{n+m}{2}}\frac{\left(n+m\right)!}{n!}\vert n,n\rangle_{1,2}.
\end{equation} 
The normalization constant is of the form $\textbf{N}^{-}_{m}\left(\lambda\right)=\left[\left(m!\right)^{2}\lambda^{m}P_{m}\left(2\lambda+1\right)\right]^{-1/2}$, where $P_{m}$ is the $m$-th order Legendre's polynomial.
Furthermore, using squeezed transformation of mode operators $a_{j}$ (and IWOP technique \cite{Fan:2003}) it is possible to generate the state in eq $(\ref{SPATSV})$ by applying squeezing operator to a $m+1$ components superposition of photon  number states, $\vert\Theta^{m}(\lambda,\chi)\rangle_{1,2}$, as it follows:

\begin{align}\label{sup1}
\vert\Psi^{(m)}_{SPATSV}\rangle_{1,2} &= \hat{S_{1,2}}\vert \Theta^{m}(\lambda,\chi)\rangle_{1,2},
\end{align}
where
\begin{align}\label{sup2}
\vert \Theta^{m}(\lambda,\chi)\rangle_{1,2}&= \sum_{k=0}^{m}C^{m}_{k}(\lambda,\chi)\vert k,k\rangle_{1,2},
\end{align}
and
\begin{align}\label{sup3}
C^{m}_{k}(\lambda,\chi)= e^{i\chi m}  \sqrt{\frac{ (1+\lambda)^{m} }{ P_{m} (2\lambda+1)  }} \times\\ \nonumber
 \times e^{i\chi k}{m\choose k}  \left(\sqrt{\frac{\lambda}{\lambda+1}}\right)^{k}
\end{align}

with $\sum_{k}\left | C^{m}_{k}(\lambda,\chi) \right |^{2}=1$.  Interestingly, $\vert \Theta^{m}(\lambda,\chi)\rangle_{1,2}$ is similar to a truncated TSV up to the components with $k\leq m$. As it can be seen by a careful inspection of $C^{m}_{k}(\lambda,\chi)$, they differs only by a binomial coefficient and a normalization factor.
For $m=0$ the state $\vert\Psi^{(0)}_{SPATSV}\rangle_{1,2}$  coincides obviously with TSV.  For $m=1$, namely one photon subtraction, the corresponding normalized two components photon number superposition state is 
\begin{equation}\label{pho_m=1}
\vert\Theta^{1}(\lambda,\chi)\rangle_{1,2}=\frac{1}{\sqrt{2\lambda+1}}\left(\sqrt{1+\lambda}\vert 0,0\rangle+e^{i\chi}\sqrt{\lambda}\vert 1,1\rangle\right).
\end{equation}
This superposition state is entangled for non-zero value of $\lambda$ and resembles to the state \cite{Dell'Anno:2007}, and for $\lambda>>1$ it becomes asymptotically a maximally entangled state. 
In general, Eq.s (\ref{sup1}),(\ref{sup2}) and (\ref{sup3}) suggest that a SPATSV state can be generated by seeding the input modes of a non-linear two-mode-squeezing interaction by an opportune superposition state in the photon number basis, see Fig. \ref{generation} b. This, represent an alternative way to generate the photon subtracted states in contrast to common approach depicted Fig. \ref{generation} a, consisting in a post selection of the state, conditioned to double click events at the detectors placed in the two arms, experimentally realized through unbalanced beam splitter (BS). 

\subsection{Squeezing properties and photon statistics of the SPATSV state}\label{Squeezing and photon statistics of the SPATSV state}

\begin{figure}[ttb]
\centering
\includegraphics[width=7cm,height=10cm]{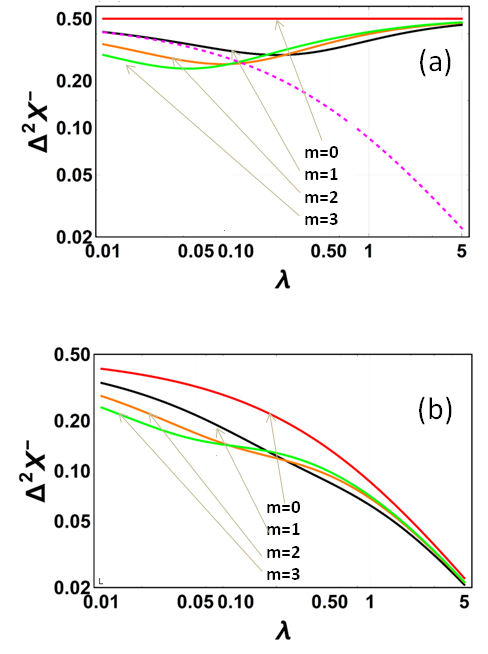}
\caption{(Color online) Non-classical amplitude quadrature correlation (0.5 is the classical bound) versus $\lambda$ for different number of photon subtraction $m$: $m=0$ (solid red line), $m=1$ (solid black line), $m=2$ ( solid orange line), and $m=3$ (solid green line). We lavel a) for superposition state $\vert \Theta^{m}(\lambda,\chi)\rangle_{1,2}$, except the dotted curve which corresponds to TSV and b) for the photon subtracted SPATSV state. }\label{squeezing}
\end{figure}

Analogously to what has been already discussed in Sec. \ref{ PASSVS}  for single mode states \cite{Wodkiewicz:1987,figurny:1993}, also two mode superposition states $\vert\Theta(\lambda,\chi)\rangle_{m}$ are squeezed in the quadrature difference even though they do not minimize the uncertainty principle.  The maximum squeezing in reached for 

\begin{equation}\label{SPATSV squeezing}
\hat{X}_{\chi}^{-}=\hat{X}_{1, \chi}-\hat{X}_{2, \chi}
\end{equation}

where $\hat{X}_{j,\chi}=(\hat{a}_{j}e^{-i\chi}+\hat{a}^{\dagger}_{j}e^{i\chi})/\sqrt{2}$ is the quadratures of the $j$-th individual input modes. This is reported in Fig. \ref{squeezing} (a). Note that, for $m\geq1$ non-classical correlations are always present, becoming stronger at the increasing of $m$ in the region of small $\lambda$. When $m\geq2$ the squeezing level overcomes the TSV limit (dotted purple line). The quadrature noise behavior of the seeding state has a direct effect on the squeezing properties of the SPATSV as shown in Fig. \ref{squeezing}(b), basically leading to a further noise reduction especially for $\lambda<1$ with respect to the TSV state. 
This effect is not trivially related to an energy shift and it can bring beneficial when using SPATSV state for specific interferometric schemes, as studied in Sec. \ref{correlated phase estimation}.

\begin{figure}[thb]
	\centering
	\includegraphics[width=5cm,height=8cm]{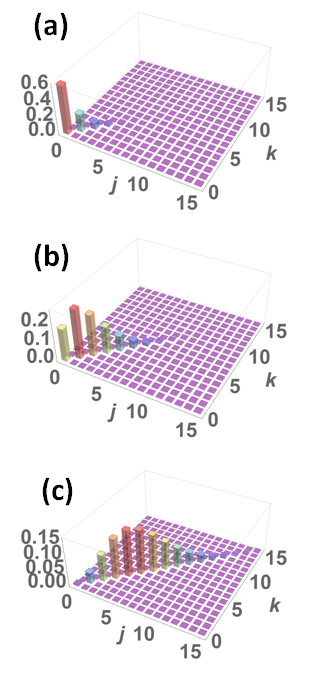}
	\caption{(Color online) 3D plots showing joint photon number distribution in SPASTV state.  $j$ and $k$ are the photon number in the mode "1" and "2", respectively. The parameters value chosen are  $ \lambda=0.6$, (a) $m=0$, (b) $m=1$, (c) $m=3$.}
	\label{pnd}
\end{figure}

\begin{figure}[thb]
	\centering
	\includegraphics[width=7 cm]{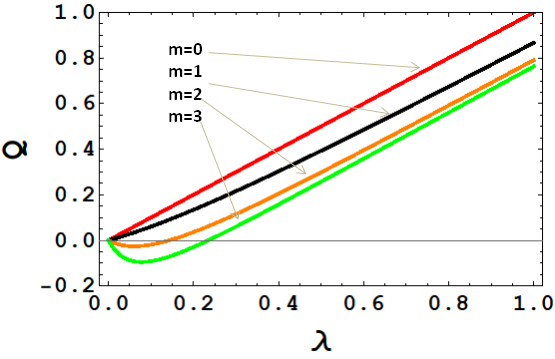}
	\caption{(Color online) Mandels $Q$ of the SPATSV state in function of the mean photon number per mode ($\lambda$) for different number of photon subtraction $m$: $m=0$ (solid red line), $m=1$ (solid black line), $m=2$ ( solid orange line), and $m=3$ (solid green line). Here, we have chosen $\eta=0.98$}\label{MandelQa}
\end{figure}
Aside quadrature squeezing, other statistical properties of the field can be improved in terms of noise reduction and turned to non-classical regime by the photon subtraction operation. Indeed, by post-selecting on the components of the state with at least one photon, induces a shrinking of the photon-number distribution because of the elimination of the vacuum-component. On one side it leads to a shift to an higher value of the mean photon number per mode $\mathcal{L}_{m}(\lambda)\geq\lambda$, as showed in Fig. \ref{pnd}. On the other side it induces a sub-shot noise behavior in each of the two modes.

Non-classicality in the photon number statistics is usually described by the  Mandels $Q$ parameter \cite{Mandel:1995}:
\begin{equation}
Q=\frac{Var(\hat{N})-\langle \hat{N}\rangle}{\langle \hat{N}\rangle},
\end{equation}
where $\hat{N}= \hat{a}^{\dagger } \hat{a}$ is the photon number operator. For classical light  Mandel parameter is bouded by $Q\geq 0$. It is worth noting  that the individual modes of TSV have thermal statistics but, once we apply the subtraction operation with $m>1$ they become non classical for low mean photons number, as evident from the negative value of Mandel's parameter reported in Fig. \ref{MandelQa}. This non-classical behavior induced by photon subtraction  is clearly not related to an energy shift of the single mode (which would conserve a thermal statistics), rather it is a more fundamental uncertainty reduction of the photon number distribution.

\subsection{Correlated phase estimation with SPASTV states}\label{correlated phase estimation}

\begin{figure}[thb]
	\centering
	\includegraphics[width=9 cm]{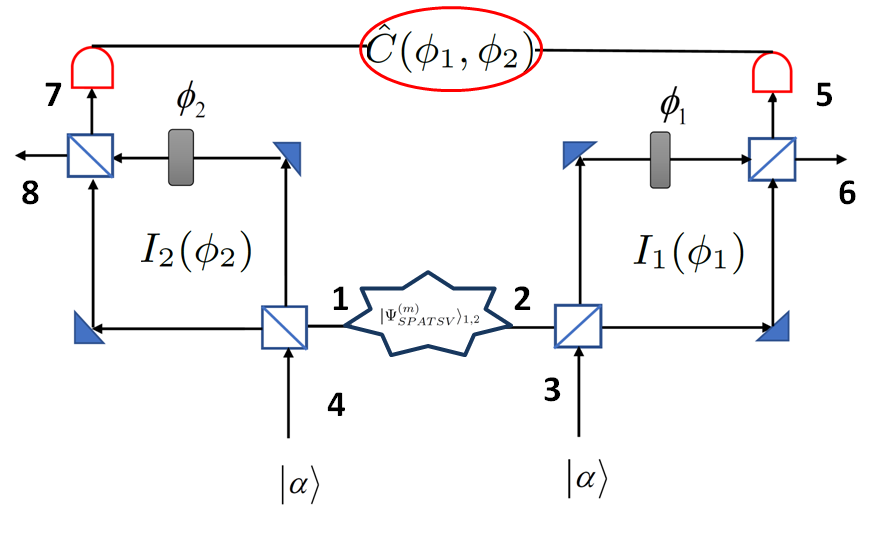}
	\caption{Correlated interferometric scheme: The modes of the bipartite input state $\vert\psi\rangle$ are mixed with two identical coherent states $\vert\alpha\rangle=\vert\mu e^{i\psi}\rangle$ in two interferometers $I_{1}(\phi_{1})$ and $I_{2}(\phi_{2})$. A joint detection is performed and the observable $\hat{C}(\phi_{1},\phi_{2})$ is measured.The losses are accounted by considering two identical detectors in both channels with the same quantum efficiency, i.e, $\eta_{5}=\eta_{7}=\eta$}\label{sketch}
\end{figure}
\par 

The interferometry system we will consider in this section is presented in Fig.\ref{sketch}. It is composed by two linear interferometers, for instance a pair of MZIs in the figure, whose photo-currents at the read-out ports are jointly measured. This is an elegant and powerful scheme in the detection of extremely faint phase signals whose magnitude can be much smaller than other sources of noise, including the shot noise. The advantage of this scheme comes from the fact that the same signal shared by the two interferometers, even if hidden in the noise in the single device, can emerges by correlating their outputs. This strategy has been already considered in several highly demanding applications, in general related to the research of stochastic fundamental backgrounds, such as gravitational wave background  \cite{Nishizawa:2008, Abadie:2012,Chou:2017a,Chou:2017} and quantum gravity effects at the Plank scale \cite{Hogan:2012,Chou:2016}. 

The advantage of using quantum state of light in such a configuration has been analyzed in ref. \cite{Berchera:2013,noi:2015}. It has been shown that injection of quantum state of light in the classically unused input ports (labeled as 1 and 2 in Fig. \ref{sketch}), either as two independent squeezed states or as a TSV allow to achieve better sensitivities. In case of TSV, for specific working conditions, i.e. very close to the dark fringes and for high quantum efficiency, the quantum advantage is dramatic even with respect to the double squeezing. Here our purpose is to investigate if and to what extent photon subtracted TVS allows to obtain better performance in virtue of their improved non-classical properties discussed in Sec. \ref{Squeezing and photon statistics of the SPATSV state}. 

\subsubsection{Noise reduction factor at the read-out ports}\label{Noise reduction factor at the read-out ports}

Let start considering the correlation properties of the read-out signals at the output ports (labeled as 5 and 7 in Fig. \ref{sketch}). In particular we are interested into photo-current subtraction, proportional to the photon number difference  $\hat{N}_{5}-\hat{N}_{7}$. 

\begin{figure}[thb]
	\centering
	\includegraphics[width=6.8 cm]{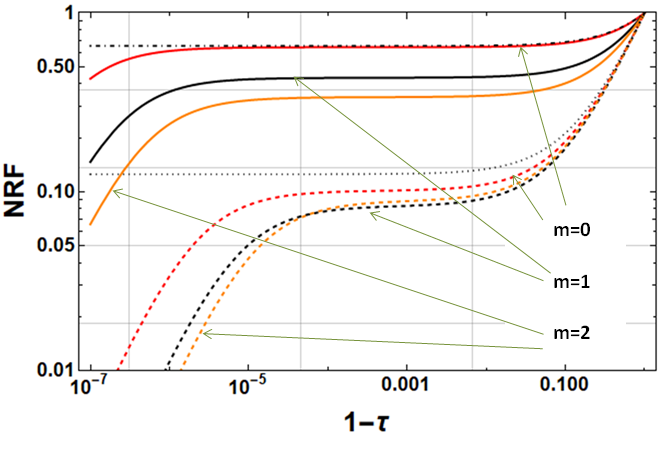}
	\caption{(Color online) Noise reduction factor at the output ports 5 and 7 of the interferometers in function of the transmittance parameter $1-\tau=\sin^{2}{\phi/2}$ for different number of photon subtraction $m$: $m=0$ (red line), $m=1$ (black line), and $m=2$ (orange line). Solid thick lines are for $\lambda=0.05$, dashed lines stand for $\lambda=2$. Asymptotic limits for $\lambda\gg1$ and for $\lambda\ll1$ ($m=0$) are the dotted and the dot-dashed lines, respectively. The other parameters are:  $\eta=1$, $\psi=\pi/2$, $\mu=10^{6}$.}\label{NRFP}
\end{figure}

Here we consider the noise reduction factor, a standard measure of non-classical correlation for a bipartite state defined as \cite{Brida:2009}
\begin{equation}\label{nrf}
\textrm{NRF}=\frac{\langle\Delta^{2}\left(\hat{N}_{5}-\hat{N}_{7}\right)\rangle}{\langle \hat{N}_{5}\rangle+\langle \hat{N}_{7}\rangle}
\end{equation}
The numerator is the variance of photon number difference and the denominator represents the standard quantum limit. Thus, $\textrm{NRF}<1$ indicates non-classical  correlation. It is convenient to introduce the factor $\tau=\cos^{2} (\phi/2)$, representing the fraction of the power at the input port 1(2), transmitted to the read-out port 5(7). Consequently, $1-\tau$ is the equivalent loss experienced by the quantum modes due to the interference fringe position and, at the same time,  the fraction of coherent power injected in ports 3(4) and transmitted to the output port 5(7). The $\textrm{NRF}$ has been evaluate analytically, and reported in Fig. \ref{NRFP} as a function of $1-\tau$. In order to analyze its behavior let us distinguish between two regimes. When the output signal is dominated by photon coming from coherent beam, i.e $\mathcal{L}_{m}(\lambda)\tau\ll\mu(1-\tau)$ (we remind $\mathcal{L}_{m}(\lambda)$ is the SPATSV mean photon number), each interferometer acts similarly to a homodyne detectors. This represents the typical working condition, since usually the coherent beam is order of magnitude brighter than the quantum light, and corresponds roughly to the region $1-\tau> 10^{-4}$ in Fig. \ref{NRFP}. In this case, the difference in the photon number at the output ports 5 and 7 becomes approximatively proportional to the difference between the quadrature of the input modes at the ports 1 and 2:

\begin{equation}\label{quadrature}
 \hat{N}_{5}-\hat{N}_{7}\propto\sqrt{\frac{\mu}{2}}\sin (\phi)\hat{X}^{-}_{\theta=\psi+\pi/2}
\end{equation}
where $\psi$ can be chosen to match the angle of the squeezed quadrature difference in Eq. (\ref{SPATSV squeezing}), in particular $\psi=\chi-\pi/2$. Therefore, the non-classical correlation of the input state reported in Fig. \ref{squeezing} b, immediately traduces in the non-classical properties of the NRF. 
In particular, for $\lambda\gg1$ the $\textrm{NRF}$ is well approximated by $\textrm{NRF}_{m}(\lambda)\approx1-\tau+\tau/(4\lambda)$ (dashed line in Fig. \ref{NRFP}) for all number of subtracted photon $m$ (for $1-\tau> 10^{-4}$ ); Also, for $\lambda\ll1$, i.e. when quadrature squeezing of the SPATSV states increases with the number of subtracted photons, the NRFs follows the same behaviour, demonstrating the advantage in using photon subtracted states, as clearly shown in Fig. \ref{NRFP}. Analytically, it can be found that in this asymptotic limit of $\lambda\ll1$ the expression of the NFR at the different orders of photon subtraction can be approximated as:
\begin{equation}\label{nrf0}
\textrm{NRF}_{m=0}\approx 1-2\tau\left(\sqrt{\lambda }- \lambda \right) 
\end{equation}
\begin{equation}\label{nrf1}
\textrm{NRF}_{m=1}\approx 1-4\tau\left(\sqrt{\lambda }- 2 \lambda \right) 
\end{equation}
\begin{equation}\label{nrf2}
\textrm{NRF}_{m=2}\approx 1-6\tau\left( \sqrt{\lambda }- 3 \lambda \right),
\end{equation}
where the first of these equation is reported in Fig. \ref{NRFP} as the dotted-dashed line.

In the opposite scenario, when the coherent beam does not contribute significantly to the outputs and the two interferometers, i.e.  $\mathcal{L}_{m}\tau\gg\mu(1-\tau)$, the interferometers can be seen as attenuators with transmission $\tau$  of the input state. The photon number entanglement between the two modes of the SPATSV input state are then preserved at the output ports for $\tau\sim1$. Indeed, in the ideal case of $\phi=0 (\tau=1)$ and unit detection efficiency, the photon number correlation is perfect, independently on the energy $\lambda$ of the input quantum state. This explains the sudden dropping down of the NRF observed in Fig. \ref{NRFP} for $1-\tau< 10^{-4}$. However, the condition $\mathcal{L}_{m}\tau\gg\mu(1-\tau)$ is reached for smaller value of $\tau$ (higher value of $\phi$) when the input energy $\mathcal{L}_{m}(\lambda)$ is larger. So, reminding that $\mathcal{L}_{m+1}(\lambda)>\mathcal{L}_{m}(\lambda)$, if the energy $\lambda$ of the TSV before the photon subtraction is fixed, subtracting more photons makes easier to reach the region in which entanglement determine a dramatic reduction of the uncertainty.  
In the next section we shall show that the characteristics of the $\textrm{NRF}$ are strictly related with the sensitivity of the double interferometric set up.

\par

% Exploiting the symmetrical properties of individual modes, we show that NRF can be expressed in another interesting form in terms of Mandels $Q$ and correlation coefficient $J=Cov(N_{5},N_{7})/\sqrt{\Delta^{2}N_{5}\Delta^{2}N_{7}}$ as follows.
%\begin{equation}\label{nrfq}
%NRF=(1+Q)(1-J).
%\end{equation}
%$J$ determines the mode correlation, lying in the range ($-1<J<1$). Thus, the non-classical correlation is simultaneously affected by the individual mode statistics and correlation. Incorporating the scheme of the quantum model \cite{noi:2015}, i.e $\tau_{1}=\tau_{2}=\tau$), the NRF can be evaluated from eq $(\ref{nrf})$. 

%This is related to the fact that $Q$ is negative in the region of $0<\lambda<0.2$ which in turn gives the correlation advantage as evident from eq $(\ref{nrfq})$. 

\subsubsection{Phase correlation estimation}

In setup of Fig. \ref{sketch} rather than the magnitude of phase noise in the single MZI, the quantity under estimation is the covariance between the phase fluctuations in two interferometers. This estimate can be somehow related to a joint measurement of the read-out signals $N_{5}$ and $N_{7}$. In the limit of faint signal, any joint observable $\widehat{C}(\phi_{1},\phi_{2})=\widehat{C}(N_{5}(\phi_{1}),N_{7}(\phi_{2}))$ with local non-null double partial derivative $\partial_{\phi_{1},\phi_{2}}^{2} C(\phi_{1}, \phi_{2})$, can be exploited for a phase-noise covariance estimation \cite{noi:2015}. Here, the goal is to investigate whether SPATSV can lead to some sensitivity advantage with respect to TSV state in that scheme. The uncertainty in the phase covariance measurement is \cite{Berchera:2013}
\begin{equation} \label{U0}
\mathcal{U} = \frac{\sqrt{2\, \mathrm{Var}\left[ \widehat{C}(\phi_{1}, \phi_{2}) \right]}}{\left|
	\partial_{\phi_{1},\phi_{2}}^{2} C(\phi_{1}, \phi_{2}) \right|}.
\end{equation}

A good choice for the joint measurement operator is $\widehat{C}(\phi_{1},\phi_{2})=\left(N_{5}(\phi_{1})-N_{7}(\phi_{2})\right)^{2}= N_{5}^{2}+N_{7}^{2}-2N_{5}N_{7}$. On one side, according to the results on the NRF discussed in Sec. \ref{Noise reduction factor at the read-out ports},  it  has fluctuation below the classical limit.  On the other, it satisfies the condition $\partial_{\phi_{1},\phi_{2}}^{2} C(\phi_{1}, \phi_{2})\neq0$.
The classical bound, obtained with coherent states at ports "3" and "4" and vacuum at the ports "1" and "2", is given by $\mathcal{U}_{cl}=\sqrt{2}\left(\eta\mu\cos^{2} [\phi/2]\right)^{-1}$ \cite{noi:2015}, where we have introduced the detection efficiency $\eta$, assumed to be equal in the two channels. Hereinafter, we present the uncertainties $U_{m}$ for $m$-th SPATSV state, as normalized to the coherent classical limit, namely $U_{m}=\mathcal{U}_{m}/\mathcal{U}_{cl}$. 

\begin{figure}[thb]
	\centering
	\includegraphics[width=7 cm]{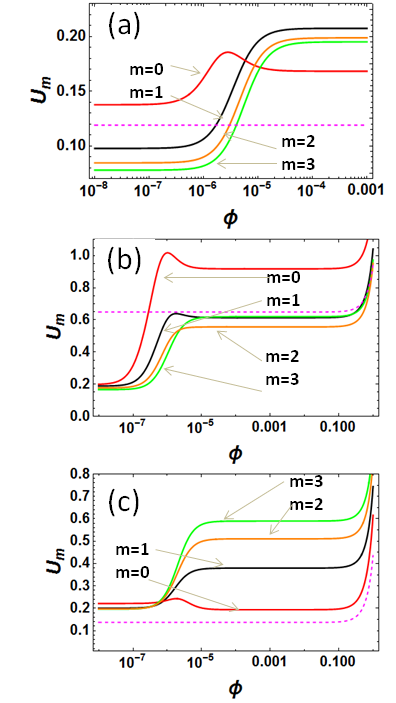}
	\caption{(Color online) Normalized uncertainty as a function of $\phi$ with $\mu=10^{12}$ for different number of photon subtraction $m$: $m=0$ (solid red line), $m=1$ (solid black line), $m=2$ ( solid orange line), and $m=3$ (solid green line). Dashed purple line represents two independent squeezed state: (a.) for $\lambda=2$, $\eta=0.98$ (b.) $\lambda=0.05$, $\eta=0.98$ (c.) energy balancing scenario for $\lambda=2$, $\eta=0.96$. }\label{uncvsphi}
\end{figure}

Analytical results of uncertainties as a function of the working central phase $\phi_{1}=\phi_{2}=\phi$ are plotted in Fig.\ref{uncvsphi}. Similarly to the case of the NRF analyzed in the previous section, one can distinguish two different regions; one laying roughly in the range $10^{-5}<\phi<\pi$ (showed up to $\phi\approx10^{-3}$ in figure) and the other for smaller value of phase, $\phi<<10^{-6}$, separated by a short transient. The range $10^{-5}<\phi<\pi$ corresponds to the situation in which the mean number of coherent photon at the read-out ports is much larger than the transmitted SPATSV photons ( i.e. $\mathcal{L}_{m}(\lambda)\tau\ll\mu(1-\tau)$). In this case, the quadrature non-classical correlation of the input modes are responsible for the read-out signal correlation. To provide compact expressions we have reported analytical results in relevant regimes.  In the limit of high coherent power, $\mu\gg1$, and low squeezing, $\lambda\ll1$, one gets:

\begin{equation}\label{unc0}
U_{m=0}\approx \sqrt{2} \left[1-\tau \eta \left(2   \sqrt{\lambda }-2  \lambda \right)\right]
\end{equation}
\begin{equation}\label{unc1}
U_{m=1} \approx \sqrt{2} \left[ 1-\tau \eta \left(4  \sqrt{\lambda }+\frac{1}{2}  \lambda  (3 \eta \tau
   -16)\right)\right]
\end{equation}
\begin{equation}\label{unc2}
U_{m=2} \approx \sqrt{2} \left[1-\tau \eta \left(6   \sqrt{\lambda }+\frac{9}{2}  \lambda  (\eta \tau
   -4)\right)\right]
\end{equation}
\begin{equation}\label{unc3}
U_{m=3} \approx \sqrt{2} \left[1-\tau \eta \left(8   \sqrt{\lambda }+ \lambda  (9 \eta \tau
   -32)\right)\right].
\end{equation}

Note that these expressions follow the $\textrm{NRF}$ behavior reported in Eq.s (\ref{nrf0},\ref{nrf1},\ref{nrf2}) up to the terms in $\sqrt{\lambda }$. 
It comes out that asymptotically, for $\lambda\ll1$, there is an advantage which increases with the number of photon subtracted $m$. However, when the asymptotic condition is not fully fulfilled, i.e. for finite values of the SPATSV energy $\mathcal{L}_{m}(\lambda)$, it can happens that at higher values of $m$ do not always correspond lower uncertainties, as  reported in Fig. \ref{uncvsphi}b in the range $\phi >10^{-5}$, e.g. for $\lambda=0.05$.
Moreover, Eq.s (\ref{unc0}-\ref{unc3}) show that the detection efficiency $\eta$ plays the same role as the interferometer transmission $\tau$  and both of them should be high enough to ensure a significant quantum advantage.

In the case of strong squeezing, $\mu\gg\lambda \gg1$, provided the condition $\mathcal{L}_{m}(\lambda)\tau\ll\mu(1-\tau)$ is still fulfilled, it turns out that the $U_{m}$'s respective expressions for different values of $m$ do not differ much, in fact we have:

\begin{equation}\label{unc4}
U_{m=0,1,2,3} \approx \sqrt{2}\left(1-\tau\eta-\frac{\tau\eta}{4\lambda}\right)
\end{equation}

\par
Also in this case, some deviation from the asymptotic behavior can emerge when finite values of the parameters are considered. For instance the case $\lambda=2$ is reported Fig. \ref{uncvsphi}a.

The opposite situation, when number of SPATSV photons are dominant with respect to the coherent ones at the read-out ports 5 and 7 ($\mathcal{L}_{m}\tau\gg\mu(1-\tau)$), corresponds in Fig.\ref{uncvsphi} to the range $\phi<<10^{-6}$.
Perfect photon number correlation of the SPATSV entangled state at the input ports 2 and 3 are preserved between $N_{5}$ and $N_{7}$. For $\mu\gg1$ (and $\phi\rightarrow 0$ ) we obtaining the following asymptotic behavior:
   
\begin{equation}\label{unc5}
U_{m=0,1,2,3}\approx \sqrt{2}\sqrt{(1- \eta)/\eta }  \qquad \lambda\ll1
\end{equation}

\begin{eqnarray}\label{unc6}
U_{m=0}&\approx2& \sqrt{5} (1-\eta )\qquad \lambda\gg1\nonumber\\
U_{m=1}&\approx2& \sqrt{3} (1-\eta )\nonumber\\
U_{m=2}&\approx2 &\sqrt{13/5} (1-\eta ))\nonumber\\
U_{m=3}&\approx2& \sqrt{17/7} (1-\eta ).
\end{eqnarray}

Eq.s (\ref{unc5}, \ref{unc6}) shows that in this regime of perfect photon number correlations the uncertainty reduction is mainly limited by the detection efficiency. This means that there exists always a value of the efficiency high enough to make this regime more advantageous with respect to the one  exploiting quadrature correlation. For example in Fig.\ref{uncvsphi}, $\eta=0.98$ guarantees a stronger advantage for $\phi<<10^{-6}$. Only or $\lambda\gg1$, the uncertainty reduction depends on the number of subtracted photons $m$. In addition, photon subtraction brings a further improvement: the increasing of the energy $\mathcal{L}_{m}$ with $m$ (at fixed $\lambda$) extends the range of validity of photon number correlation advantage, $\mathcal{L}_{m}\tau\gg\mu(1-\tau)$, towards higher values of $\phi$, as shown in Fig.\ref{uncvsphi}.

\begin{figure}[thb]
\centering
\includegraphics[width=7 cm]{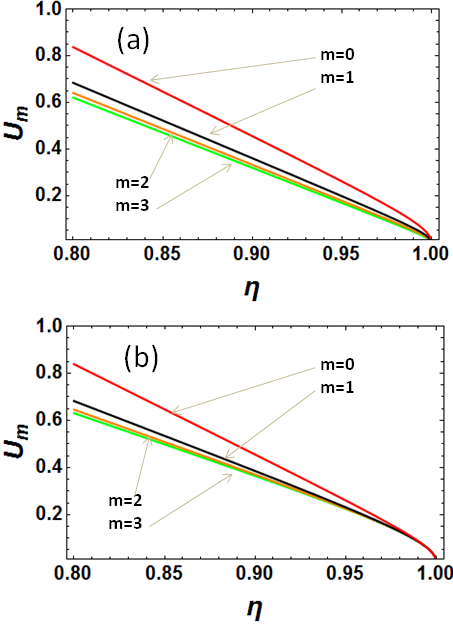}
\caption{(Color online) Normalized uncertainty versus the detection efficiency $\eta$ with $\mu=10^{12}$, $\lambda=2$, and $\phi=10^{-8}$ for different number of photon subtraction $m$: $m=0$ (solid red line), $m=1$ (solid black line), $m=2$ ( solid orange line), and $m=3$ (solid green line), (a) without energy balancing (b) balanced energy condition.}\label{unc_eta}
\end{figure}

It is relevant to understand whether the uncertainty reduction observed for SPATSV states can be explained only in terms of the mean energy increasing due to the photon subtraction operation, or the advantage comes from other properties of these states. Also in this case, we consider the energy balancing approach as in Sec. \ref{Single phase estimation with PASSV states}, where the energies of two mode photon subtracted states  ($m=0,1,2,3$) are made equivalent to the energy of TSV, and we observe that the uncertainty reduction advantage in the high detection efficiency cases (Fig. \ref{uncvsphi}a-b) almost disappears (see Fig.  \ref{uncvsphi}c). 
However, in case of realistic value of the detection efficiency and optical losses, we observe that the improvement of uncertainty reduction is still present (Fig. \ref{unc_eta}). For instance, in this scenario SPATSV with $m=3$ present around 26 \% of uncertainty reduction advantage compared to TSV at a detection efficiency of 0.8. Thus, the uncertainty reduction obtained with SPATSV states is, in general, not only due to the energy shifts, but it also comes from the enhancement in mode correlation and statistics. 

\section{Summary and conclusions}\label{Summary and conclusions}
We have studied in detail multi photon subtracted one- and and two-mode squeezed vacuum state, in relation to phase estimation in both single and correlated interferometry. 
%We have obtained new compact form expressions which show photon  subtraction is an operation that results equivalent to the squeezing of a certain finite dimension superposition states in the photon number basis. Quadrature squeezing is always associated with superposition states, and more is the component of superposition, the stronger is the quadrature squeezing. 
The squeezing of the single mode PASSV state not necessarily improves with the number of subtracted photons. In the case of odd number of photon subtraction, it is definitely worse than SSV, while for even photon subtraction, it is better than the SSV only for relatively small brightness. The phase estimation uncertainty  in a single interferometer reflects this behavior, as expected. Moreover, by comparing the phase sensitivity after re-adjusting the energy of the  PASSVs to match the one of SSV, the advantage of the photon subtraction disappears, at least at the optimal working point of $\phi=\pi/2$. For other values of the central operating phase we have different behavior and in some cases, as shown in Fig. \ref{anyangle}, the advantage of photon subtraction is preserved even when energies are balanced. 

In terms of QFI, we have found improvements with the number of photon subtraction, but for energy balancing condition this advantage disappears. However, Heisenberg limit can be reached for an asymptotically large number of photons in a lossless interferometer. 

We also investigated SPATSV for correlated interferometry \cite{Berchera:2013, noi:2015}. Usually such SPATSV states are generated by probabilistic events with low success rate. We showed analytically how symmetric photon subtraction from two mode squeezed vacuum is equivalent to the squeezing of a finite component superposition state, suggesting an alternative way for the generation of SPATSV states. We found that, these SPATSV states always show quadrature squeezing and their strength increases with the number of symmetrical photon subtraction for small energy of the state. Various statistical properties including photon number distribution, Mandel's $Q$ function, and noise reduction factor shows higher non-classical signature of SPATSV with respect to TSV suggesting its potential advantages in correlated phase estimation. In fact, concerning the phase correlations measurement among two interferometers, we observed that SPATSV are able to achieve a smaller uncertainty than TSV for an operation point close to the dark fringe ($\phi \approx 0$). In the low losses scenario, SPATSV apparently provide substantial advantage in uncertainty reduction  with respect to TVS state, but this is essentially explained by the energy increasing of the states due to the photon-subtraction. In fact, re-normalizing the energies of the SPATSV to the one of the initial TSV state, the uncertainty reduction is lost. However, SPATSV conserve an advantage of about 30$\%$ with respect to TVS in the high loss scenario and this can be attributed to their improved statistical properties.
\section{ACKNOWLEDGMENTS}
We acknowledge the funding from the European Unio{n}'s Horizon 2020 research and innovation programme under  Grant Agreement  No. 862644 (QUARTET) and from  the European Unio{n}'s Horizon 2020 research and innovation programme and EMPIR Participating States in the context of the project 17FUN01 (BeCOMe).


\begin{thebibliography}{100}
\bibitem{Genoni:2010} M.G. Genoni and M.G.A. Paris, PRA 82, 052341 (2010)
\bibitem{Olivares:2003} Olivares et. al, "Teleportation improvement by inconclusive photon subtraction", Phys. Rev. A {\bf 67}, 032314  (2003).
\bibitem{Optarny:2000} T. Opatrny, G. Kurizki, D.-G. Welsch, "Improvement on teleportation of continuous variables by photon subtraction
via conditional measurement", Phys. Rev. A {\bf 61}, 032302  (2000).
\bibitem{Cochrane:2002} P. T. Cochrane, T. C. Ralph, and G. J. Milburn, "Teleportation improvement by conditional measurements on the two-mode squeezed vacuum", Phys. Rev. A {\bf 65}, 062306  (2002).
\bibitem{Dell'Anno:2010} F. Dell?Anno, S. De Siena, and F. Illuminati Phys. Rev. A 81, 012333 (2010).

\bibitem{Borelli:2016} L.F.M. Borelli, L.S. Aguiar, J.A. Roversi and A. Vidiella-Barranco, "Quantum Key Distribution using Continuous-variable
non-Gaussian States",Quantum Inf Process  {\bf 15}, 893?904 (2016).

\bibitem{Liao:2018} Qin Liao et. al, "Long-distance continuous-variable quantum key distribution using
non-Gaussian state-discrimination detection", New. J. Phys {\bf 20}, 023015  (2018).

\bibitem{Cerf:2005} N. J. Cerf, O. Kruger, P. Navez, R. F.Werner, and M. M.Wolf, "Non-Gaussian Cloning of Quantum Coherent States is Optimal", Phys. Rev. Lett. {\bf 95}, 070501  (2005).

\bibitem{Eisert:2002} J. Eisert, S. Scheel, and M. B. Plenio, Phys. Rev. Lett. {\bf 89}, 137903  (2002).

\bibitem{Cirac:2002} G. Giedke and J. I. Cirac, Phys. A {\bf 89}, 032316 (2002).

\bibitem{Fiurasek:2002} J. Fiurasek, Phys. Lett. {\bf 89}, 137904 (2002).

\bibitem{Niset:2009} J. Niset, J. Fiurasek, and N. J. Cerf, Phys. Lett. {\bf 102}, 120501 (2009).
\bibitem{Ferreyrol:2010} F. Ferreyrol, M. Barbieri, R. Blandino, S. Fossier, R. Tualle-
Brouri, and P. Grangier, Phys. Lett. {\bf 104}, 123603 (2010).

\bibitem{Xiang:2010} G. Y. Xiang, T. C. Ralph, A. P. Lund, N.Walk, and G. J. Pryde, Nature Photonics. {\bf 4}, 316 (2010).

\bibitem{Banaszek:1998} K. Banaszek and K. W´odkiewicz, Phys. A. {\bf 58}, 4345 (1998).

\bibitem{Nha:2004} H. Nha, and H. J. Carmichael, Phys. Lett. {\bf 93}, 020401 (2004).

\bibitem{Takagi:2018} R. Takagi and Q. Zhuang, Phys. A. {\bf 97}, 062337  (2018).

\bibitem{Albarelli:2018} F. Albarelli et. al., Phys. A. {\bf 98}, 052350  (2018).

\bibitem{Genoni:2009} M. G. Genoni, C. Invernizzi, and M. G. A. Paris, "Enhancement of parameter estimation by Kerr interaction", Phys. A. {\bf 80}, 033842 (2009).

\bibitem{raul:2012} R. Carranza and C. C. Gerry, "Photon-subtracted two-mode squeezed vacuum states and applications to quantum-optical interferometry", J. Opt. Soc. Am. B {\bf 29}, 2581 (2012).

\bibitem{kim:2008} MS Kim,"Recent developments in photon-level operations on travelling light fields", J. Phys. B: At. Mol. Opt. Phys.{\bf 41}, 133001 (2008).

\bibitem{kim1:2008} M. Kim,"The quantum mechanics of photon addition and subtraction", Optoelectronics Comm. {\bf 4},(2008).
\bibitem{Bellini:2007} V. Parigi, A. Zavatta, M. S. Kim, M. Bellini, "Probing quantum commutation rules by addition and subtraction of single photons to/from a light field", Science {\bf 317}, 1890-1892, (2007).

\bibitem{tara:1990} Agarwal, G. S. and Tara, K., "Nonclassical properties of states generated by the excitations on a coherent state", Phys. Rev. A {\bf 43}, (1991).

\bibitem{Bellini:2004} A. Zavatta, S. Viciani, M. Bellini, "Quantum-to-classical transition with single-photon-added coherent states of light", Science {\bf 306}, 660-662, (2004).

\bibitem{Bellini1:2007} A. Zavatta, V. Parigi, M. Bellini, "Experimental nonclassicality of single-photon-added thermal light states", Phys. Rev. A {\bf 75}, 052106 (2007).

\bibitem{Carlos:2012} Navarrete-Benlloch, Carlos and Garc\'{\i}a-Patr\'on, Ra\'ul and Shapiro, Jeffrey H. and Cerf, Nicolas J., "Enhancing quantum entanglement by photon addition and subtraction", Phys. Rev. A {\bf 86}, 012328 (2012).

\bibitem{Tim:2013} Bartley, Tim J. and Crowley, Philip J. D. and Datta, Animesh and Nunn, Joshua and Zhang, Lijian and Walmsley, Ian, "Strategies for enhancing quantum entanglement by local photon subtraction", Phys. Rev. A {\bf 87},022313 (2013).

\bibitem{ourjoumtsev:2007} Ourjoumtsev, Alexei and Dantan, Aur\'elien and Tualle-Brouri, Rosa and Grangier, Philippe, "Increasing Entanglement between Gaussian States by Coherent Photon Subtraction", Phys. Rev. Lett.  {\bf 98},030502 (2007).

\bibitem{chekova:2016} Iskhakov, Timur Sh. and Usenko, Vladyslav C. and Filip, Radim and Chekhova, Maria V. and Leuchs, Gerd, "Low-noise macroscopic twin beams", Phys. Rev. A {\bf 93}, 043849 (2016).

\bibitem{Zhang:2014} Zhang, ShengLi and Guo, JianSheng and Bao, WanSu and Shi, JianHong and Jin, ChenHui and Zou, XuBo and Guo, GuangCan, "Quantum illumintion with photon subtracted continuous variable entanglement", Phys. Rev. A {\bf 89}, 062309 (2014).

\bibitem{Lloyd:2008} Lloyd S, "Enhanced sensitivity of photo detection via quantum illumination", Science {\bf 321} 1463 (2008).

\bibitem{Lopaeva:2013} Lopaeva E D, Ruo Berchera I, Degiovanni I P, Olivares S, Brida G, Genovese M, "Experimental realization of quantum illumination", Phys. Rev. Lett {\bf110}, 153603 (2013).

\bibitem{Birritella:2014} Richard Birrittella and Christopher C. Gerry, "Quantum optical interferometry via the mixing of coherent and photon-subtracted squeezed vacuum states of light", J. Opt. Soc. Am. B {\bf 31}, 3 (2014).

\bibitem{yoa:2016} Y. OUYANG, S. WANG, and L. ZHANG, "Quantum optical interferometry via the photon added two-mode squeezed vacuum states", J. Opt. Soc. Am. B {\bf 33}, 1373 (2016).

\bibitem{Caves:1981} Caves, Carlton M., "Quantum-mechanical noise in an interferometer", Phys. Rev. D {\bf 23}, 1708 (1981).

\bibitem{Tse:2019} M. Tse et. al., "Increasing the Astrophysical Reach of the Advanced Virgo Detector via the Application of Squeezed Vacuum States of Light
", Phys. Rev. Lett {\bf 123}, 231108 (2019).

\bibitem{Acernese:2019} F. Acernese  et. al., " Increasing the Astrophysical Reach of the Advanced Virgo Detector via the Application of Squeezed Vacuum States of Light
", Phys. Rev. Lett {\bf 123}, 231108 (2019).


\bibitem{Rafal:2013} R. Demkowicz-Dobrza´nski, K. Banaszek, and R. Schnabel, "Fundamental quantum interferometry bound for the squeezed-light-enhanced gravitational wave detector GEO 600", Phys. Rev. A {\bf 88},  041802(R) (2013).


\bibitem{olivares:2016} S. Olivares, M. Popovic, and M. G. A. Paris, "Phase estimation with squeezed single photons", Quantum Meas. Quantum Metrol. {\bf 3} 38-43 (2016).

\bibitem{Dell'Anno:2007} Dell' Anno et. al, "Continuous-variable quantum teleportation with non-Gaussian resources", Phys. A. {\bf 76}, 022301 (2007).

\bibitem{Marek:2018} P. Marek, J.Provatnzic, R.Filip, Loop-based subtraction of a single photon from a traveling beam of light, Opt. Exp. 26,  29837(2018).

\bibitem{Wodkiewicz:1987} K. Wodkiewicz, P. L. Knight, S. J. Buckle, and S. M. Barnett, "Squeezing and superposition states", Phys. Rev. A {\bf 35}, 2567 (1987).

\bibitem{figurny:1993} P. Figurny, A. Orlowski, and K. Wodkiewicz, "Squeezed fluctuations of truncated photon operators", Phys. Rev. A {\bf 47}, 5151 (1993).

\bibitem{nigam:2017} K. Thapliyal, N. Samantaray, J. Banerji and A. Pathak, "Comparison of lower- and higher-order nonclassicality in photon added and subtracted squeezed coherent states", Phys. Lett. A {\bf 381}, 37 (2017).

\bibitem{Fan:2003} Hong. Y. Fan, "Operator ordering in quantum optics theory and the development of Dirac s symbolic method", J. Opt. B: Quantum. Semiclass. Opt. {\bf 5} (2003).

\bibitem{Demkowicz:2015} R. Demkowicz-Dobrzanski, M. Jarzyna, J. Kolodynski, "Quantum Limits in Optical Interferometry", Prog. Opt. {\bf 60}, 345 (2015).

\bibitem{Pezze:2008} L. Pezze and A. Smerzi, "Mach-Zehnder Interferometry at the Heisenberg Limit with Coherent and Squeezed-Vacuum Light", Phys. Rev. Lett. {\bf 100}, 073601 (2008).

\bibitem{Mandel:1995} Mandel, L., and Wolf, E., "Optical Coherence and Quantum Optics" Cambridge (1995).

\bibitem{Nishizawa:2008} Atsushi Nishizawa, Seiji Kawamura, Tomotada Akutsu, Koji Arai, Kazuhiro Yamamoto, Daisuke Tatsumi, Erina Nishida, Masa Aki Sakagami, Takeshi Chiba, Ryuichi Takahashi, and Naoshi Sugiyama, "Optimal location of two laser-interferometric detectors for gravitational wave backgrounds at 100 MHz",
Classical and Quantum Gravity {\bf 25}, 225011 (2008).

\bibitem{Abadie:2012} J. Abadie et. al., "Upper limits on a stochastic gravitational-wave background using LIGO and Virgo interferometers at 600-1000 Hz",
 Phys. Rev. D {\bf 85}, 122001 (2012).

\bibitem{Chou:2017a} Aaron S. Chou, Henry Glass, H.Richard Gustafson, Craig J. Hogan, Brittany L. Kamai, Ohkyung Kwon, Robert Lanza, Lee McCuller, Stephan S. Meyer, Jonathan W. Richardson, Chris Stoughton, Ray Tomlin, and Rainer Weiss., "Interferometric constraints on quantum geometrical shear noise correlations",
Classical and Quantum Gravity {\bf 34}, 165005 (2017).

\bibitem{Chou:2017} Aaron S. Chou, Richard Gustafson, Craig J. Hogan, Brittany Kamai, Ohkyung Kwon, Robert Lanza, Shane L. Larson, Lee McCuller, Stephan S. Meyer, Jonathan Richardson, Chris Stoughton, Raymond Tomlin, and Rainer Weiss., "MHz gravitational wave constraints with decameter Michelson interferometers".
Phy. Rev. D {\bf 96}, 063002 (2017).


\bibitem{Hogan:2012} Craig J. Hogan., "Interferometers as probes of Planckian quantum geometry", Phy. Rev. D {\bf 85}, 064007 (2012).


\bibitem{Chou:2016} Chou, Aaron S. and Gustafson, Richard and Hogan, Craig and Kamai, Brittany and Kwon, Ohkyung and Lanza, Robert and McCuller, Lee and Meyer, Stephan S. and Richardson, Jonathan and Stoughton, Chris and Tomlin, Raymond and Waldman, Samuel and Weiss, Rainer, "First Measurements of High Frequency Cross-Spectra from a Pair of large Michelson Interferometers", Phys. Rev. Lett. {\bf 117}, 111102 (2016).

\bibitem{Berchera:2013} Ruo Berchera, I. and Degiovanni, I. P. and Olivares, S. and Genovese, M. "Quantum Light in Coupled Interferometers for Quantum Gravity Tests", Phys. Rev. Lett. {\bf 110}, 213601 (2013).

\bibitem{noi:2015} Ruo-Berchera, I. and Degiovanni, I. P. and Olivares, S. and Samantaray, N. and Traina, P. and Genovese, M., "One and two-mode squeezed light in correlated interferometry", Phys. Rev. A {\bf 92}, 053821 (2015).

\bibitem{Brida:2009} Brida et.al, "Measurement of sub-shot-noise spatial correlations without background subtraction", Phys. Rev. Lett. {\bf 102}, 213602 (2009).
\end{thebibliography}
\end{document}